\documentclass[aps,pra,reprint,showpacs]{revtex4-1}

\usepackage{bm}
\usepackage{amssymb}
\usepackage[intlimits]{amsmath}
\usepackage{graphicx}
\usepackage{xcolor}
\definecolor{olive}{rgb}{0,0.5,0}
\definecolor{maroon}{rgb}{0.5,0,0}

\newcommand{\ham}{\hat{\bm H}}

\begin{document}

\title{Quantum transport properties in Datta-Das tuned opacity spin-transistors} 

\author{R. Cuan}
\email[E-mail me at: ]{rcuan@fisica.uh.cu}
\affiliation{Facultad de F\'{\i}sica. Universidad de La Habana, C.P.10400, La Habana, Cuba.}

\author{J. J. Gonz\'{a}lez}
\affiliation{Facultad de F\'{\i}sica. Universidad de La Habana, C.P.10400, La Habana, Cuba.}

\author{L. Diago-Cisneros}
\affiliation{Facultad de F\'{\i}sica. Universidad de La Habana, C.P.10400, La Habana, Cuba.}

\date{\today}

\begin{abstract}
We studied the spin-dependent quantum transport properties using a simple modelling of a Datta-Das spin transistor.
We refine previous results by accounting the propagation medium changes of opacity felt by itinerant electrons, when the gate-voltage is switched on and modelling them  \textit{via} the transversal energy levels mismatch.
Monitoring the topological-dependent conductance, we are able to identify the device operating points.
If the incoming electrons energy approaches the biased-induced barriers height, the spin-resolved conductance oscillations become significant.
In a zero temperature picture, our computations of the spin-dependent conductance as function of the electric field at the region below the gate electrode suggest the feasibility of the  modeled device. Although we demonstrate that phase time may not be spin-resolved, our simulation allows us to evaluate the time that takes an electron to experience a spin-flip process, resulting in an order of magnitude lower than typical values of the spin relaxation times.
\end{abstract}


\maketitle 

\section{Introduction}

One of the biggest challenges in Spintronics \cite{Awschalom07,Chen14} is to achie\-ve an efficient manipulation of the spin degree of freedom at zero magnetic field. Rashba spin-orbit coupling (SOC-R) \cite{Rashba84a,Rashba84b} is among the most promising phenomena. SOC-R is a natural consequence of an asymmetry in the confining potential of carriers in low dimensional semiconductor systems, so-called structure inversion asymmetry (SIA) \cite{BookWinkler03}. To intuitively understand SOC-R some authors consider that electrons feel a potential gradient due to SIA, which Lorentz transforms into an effective magnetic field that acts over theirs spin. Nonetheless, attention most be paid since Lorentz transformation neglects the atomic cores contribution to SOC-R, felts by a Bloch electron in a solid \cite{BookWinkler03}.

The strength of SOC-R can be tuned by external electric fields \cite{Nitta09}, which imply a gate mechanism for spin-based transistors \cite{Datta90,Schliemann03,Shelykh05,Chuang15}. Datta and Das spin field effect transistor (SFET) \cite{Datta90} (Fig.~\ref{fig:DDSFET}) is a theoretical device, conceptually similar to an electro-optic modulator.  Two ferromagnets act as polarizer and analyzer. The propagation medium between them ---capable of inducing a gate-controllable net rotation of the spin orientation \textit{via} SOC-R--- is a two-dimensional electron gas (2DEG).

Spin-dependent conductance in SFET-like systems has been widely addressed \cite{Mireles01,Mireles02,Pala04,Pala04b,Yang08,Gao11,Pareek02,Wu03}. In 2001, Mireles and Kirczenow \cite{Mireles01} perform calculations of spin-dependent ballistic transport properties in quantum wires, in the presence of SOC-R. Using a tight-binding scheme, they showed that a strong SOC-R may lead to dramatic changes in the transmission of electrons. The  transport of holes in $p$-doped hybrid (magnetic and nonmagnetic) structures was considered by Pala \textit{et al.} \cite{Pala04,Pala04b}, exhibiting clear possibilities of current manipulation. The effects of the temperature \cite{Mireles02}, external magnetic fields \cite{Yang08} and the direction of magnetization at the polarizer and analyzer electrodes \cite{Yang08,Pareek02,Pala04} have also been addressed. Recently, Gao \textit{et al.} \cite{Gao11} carried out a simulation of a SFET, based on the nonequilibrium Green's function formalism self-consistently coupled with a Poisson solver to produce the device $I-V$ characteristics,  obtaining good agreement with the recent experiments. However, these studies does not allow for the modifications in the opacity of the propagation medium that itinerant electrons feel when the gate-voltage is switched, as we will discuss later.

Another issue insufficiently investigated is the spin-dependent tunneling time in SFET-like systems. As far as we know, only few studies have focused the quantum transport time under SOC-R in ferromagnetic/semicon\-duc\-tor/ferromagnetic systems. Based on the group velocity concept, Wu \textit{et al.} \cite{Wu03} showed that as the strength of the SOC-R increases, the traversal time considerably decreases and exhibits step-like behavior when the length of the channel growths. However, it should be pointed out that SOC-R appears in the system they are modelling as a consequence of an asymmetry introduced at the ferromagnetic/semiconductor interface to fix some known problems related with the efficiency of the spin-injection. Then, as the potential gradient in this case is parallel  to the transmission direction, the effective magnetic field is null. Hence, there is not spin-precession, in contrast to SFET-like systems where the potential gradient induced by the gate electrode is perpendicular to the transmission direction.

This contribution focuses on the quantum transport properties in SFET-like devices. Using a simple modelling, combined with a suitable algorithm of solution, we show that spin-depend conductance exhibits the general features reported before, but strongly modulated by dispersive effects, appearing when the SOC-R coupling parameter is tuned \textit{via} external electric fields. Although we demonstrate that the transport time ---using the phase time concept--- may not be spin-resolved, our simulation allows us to evaluate the spin-flip time, which is an important time scale for spintronics applications, together with the spin relaxation and decoherence times.

The remaining part of this paper is organized as follows: Sec.~\ref{sec:model} is devoted to describe the physical system and the theoretical formalism describing electrons in SFET-like systems. The general properties of the simulation procedure are presented in Sec.~\ref{sec:MSA}. In Sec.~\ref{sec:results} we discuss numerical results of spin-dependent conductance and phase time. Finally, in Sec.~\ref{sec:Conc}, we underline some concluding remarks.

\section{Theoretical Model}
\label{sec:model}

\begin{figure}
\centering
\includegraphics[width=0.9\linewidth]{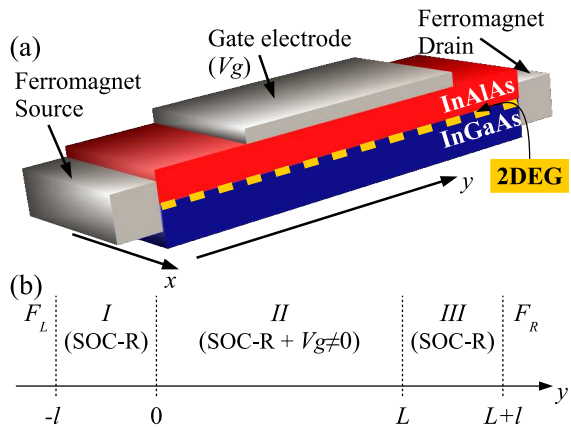}
\caption{(Color online) (a) Sketch of the Datta and Das SFET. Two ferromagnets act as analyzer and polarizer, the propagation medium ---capable of inducing a gate-controllable net rotation of the spin orientation \textit{via} SOC-R--- is a two-dimensional electron gas (2DEG). (b) The five regions we are considering: $F_L$ and $F_R$ correspond to the ferromagnets, whereas $I$, $III$ and $II$ stand for the regions free and the one right below the gate electrode, respectively.}
\label{fig:DDSFET}
\end{figure}

We simulate the device depicted in Fig.~\ref{fig:DDSFET} by a five regions system: $F_L$ and $F_R$ correspond to the ferromagnets, whereas $I$ ($III$) and $II$ stand for the free and gate-managed regions, respectively.  The 2DEG is confined at the interface of a heterostructure, InAlAs/InGaAs for instance, where SIA lead to Rashba coupling. Spin-polarized electrons are injected by an ideal ferromagnet from the region $F_L$. Another ideal ferromagnet allows identifying the spin direction of the right side outgoing electrons. By ``ideal'' we overlook the actual limitations of ferromagnets  \cite{Schmidt00,Rashba00}, fixing the initial electrons spin polarization instead. For simplicity, we choose the same effective mass along the whole system.

The [$xy$]-plane total Hamiltonian for a 2DEG confined on it, has the form \cite{Mireles01}
\begin{equation}\label{eq:HSitema}
\hat{\bm H}=\begin{bmatrix}
             \dfrac{\hbar^2}{2m^*}(\hat k_x^2 + \hat k_y^2)+E_{bs} & i\alpha_R \hat k_x+\alpha_R\hat  k_y\\
             -i\alpha_R \hat k_x+\alpha_R\hat  k_y & \dfrac{\hbar^2}{2m^*}(\hat k_x^2 + \hat k_y^2)+E_{bs}
             \end{bmatrix}.
\end{equation}

Here we have introduced an extra term $E_{bs}$, standing for the bounded states at the $z$ direction. In the infinite triangular well approximation to the heterostructure confining potential \cite{Andrada94,Cuan11}, $E_{bs}$ are given by
\begin{equation}\label{eq:AEz}
E_{bs \,n}=C_n\left(\frac{(\hbar F)^2}{2m^*}\right)^{1/3},\quad\text{with: } n=1,2,3,\dots,
\end{equation}
where $C_n$ are zeros of $A_i$ Airy function  ($C_1=2.338$, $C_2=4.088$, $C_3=5.521$, \dots), and $F=\frac{e^2n_s}{\epsilon_0\epsilon} + F_{\text{ext}}$ the electric field due to SIA. Note that $F$ involves both intrinsic ---density ($n_s$) dependent--- and external ($F_{\text{ext}}$) contributions.

The off-diagonal terms in (\ref{eq:HSitema}) correspond to the SOC-R, where $\alpha_R$ is the coupling parameter \cite{Andrada94}
\begin{equation}\label{eq:RashbaPar}
 \alpha_R(F)=F\frac{\hbar^2}{2m^*}\frac{\varDelta_{SO}}{E_g}\frac{2E_g+\varDelta_{SO}}{(E_g+\varDelta_{SO})(3E_g+2\varDelta_{SO})}.
\end{equation}
$E_g$ and $\varDelta_{SO}$ are the gap and splitt-off energies, respectively.

Because of Rashba coupling, the eigenvalue problem for the Hamiltonian (\ref{eq:HSitema}) $\ham \bm\varPsi=E\bm\varPsi$ produces two branches in the dispersion relations
\begin{equation}\label{eq:DispR}
 E_1(k)=\frac{\hbar^2}{2m^*}k^2-\alpha_R k\qquad\mbox{and}\qquad E_2(k)=\frac{\hbar^2}{2m^*}k^2+\alpha_R k,
\end{equation}
where $k=\sqrt{k_x^2+k_y^2}$, and eigenvectors
\begin{equation}\label{eq:Eigen}
 \bm\chi_{1}=\frac{1}{\sqrt{2}}\begin{bmatrix}
                 1\\
                 +ie^{i\theta}
\end{bmatrix}\qquad\mbox{and}\qquad
\bm\chi_{2}=\frac{1}{\sqrt{2}}\begin{bmatrix}
                 1\\
                 -ie^{i\theta}
\end{bmatrix},
\end{equation}
with $\theta=\arctan k_x/k_y$.

Hamiltonian (\ref{eq:HSitema}) is time-reversal invariant, therefore it can not  support spontaneous spin polarization of the electron states \cite{Mireles01}.  However, it is capable of removing the spin degeneracy for $k\neq 0$ through an effective magnetic filed \cite{Ganichev04} $\bm B_{\textrm{eff}}=\frac{1}{\mu_B}\alpha_R(-k_y,k_x,0)$, with $\mu_B$ as the Bohr magneton.

\section{Multicomponent Scattering Approach}
\label{sec:MSA}

Weakly coupled multi-mode quantum transport of electrons has been typically treated through one-dimensional single-mode approximations. The corresponding solutions, along with the envelope function approximation, are multicomponent wave functions. However, it is usual \cite{Wessel89,Kumar97} to  arbitrarily cancel all components but one in order to compute transmission coefficients, which may lead to lose valuable physical information.  Recently, an alternative formalism, named Multicomponent Scattering Approach (MSA), has been proposed (See Refs. \cite{Leo02,Leo06,Leo07,Arias-Laso12} for details). MSA takes into account ---jointly and simultaneously--- all the propagating modes, then a multicomponent description of transmission amplitudes is carried out in a natural way.

Let
\begin{equation}\label{eq:solutions}
 \bm F(y)=\sum_{j=1}^{2N}c_j\bm\chi_je^{ik_jy} = \sum_{j=1}^{2N}c_j\bm F_j(y)
\end{equation}
be  the general solution of the system described by the Hamiltonian (\ref{eq:HSitema}). To introduce the transfer matrices we define the bi-vector
\begin{equation}\label{eq:bi-vec-fd}
\bm \Psi(y)=\left(\begin{array}{c}
             \bm F(y)\\
             \bm F'(y)
            \end{array}\right),
\end{equation}
and the state vector
\begin{equation}\label{eq:state-v}
 \bm \Phi(y)=\left( \begin{array}{cc}
                \bm a & 0\\
                0 & \bm b
               \end{array}\right)
                \left( \begin{array}{c}
                 \overrightarrow{\bm\varphi}(y)\\
                 \overleftarrow{\bm \varphi}(y)
               \end{array}\right).
\end{equation}
$\bm \Psi$ include the wave vectors and their derivative. The coefficients $\bm a$ and $\bm b$ are  matrices  ($2\times 2$). $\overrightarrow{\bm\varphi}(z)$ and $\overleftarrow{\bm \varphi}(z)$ are bidimensional vectors, whose components describe a propagating or evanescent mode, according to the energy.

Based on the above definitions and taking into account that solutions (\ref{eq:solutions}) are plane-waves, it is possible to establish the crucial relation
\begin{equation}
 \bm \Psi(z)=\mathcal{\bm N} \bm \Phi(z).
\end{equation}
The matrix $\mathcal{\bm N}$ depends on the specific Hamiltonian, and is build up using the property $\left( e^{ax}\right)'=ae^{ax}$. In our particular case we have
\begin{equation}
 \mathcal{\bm N}=
 \begin{pmatrix}
  1 & 0 & 1 & 0 \\
  0 & 1 & 0 & 1 \\
  ik & 0 & -ik & 0 \\
  0 & ik & 0 & -ik
 \end{pmatrix},
\end{equation}
being $k=(1/\hbar)\sqrt{2m^{*}E}$.

The first-kind transfer matrix $\bm M_{fd}$ ($fd$ stands for function and derivative),  which relates the solutions and their derivative (\ref{eq:bi-vec-fd}) between two points of the system $y_{\mbox{\tiny{L}}}$ and $y_{\mbox{\tiny{R}}}$, is defined by
\begin{equation}\label{eq:Mfd}
 \bm \Psi(y_{\mbox{\tiny{R}}})=\bm M_{fd}(y_{\mbox{\tiny{R}}},y_{\mbox{\tiny{L}}})\bm \Psi(y_{\mbox{\tiny{L}}}).
\end{equation}
Likewise, the second-kind transfer matrix $\bm M_{sv}$ ($sv$ stands for state vector), which relates the sate vectors (\ref{eq:state-v}), is defined by
\begin{equation}\label{eq:Msv}
\bm \Phi(y_{\mbox{\tiny{R}}})=\bm M_{sv}(y_{\mbox{\tiny{R}}},y_{\mbox{\tiny{L}}})\bm \Phi(y_{\mbox{\tiny{L}}}).
\end{equation}
These matrices satisfy the relevant physical properties and symmetries of the Hamiltonian (\ref{eq:HSitema}).

Using equations (\ref{eq:bi-vec-fd})-(\ref{eq:Msv})  the transformation
 \begin{equation}
\bm{M}_{sv}(y_{\mbox{\tiny{R}}},y_{\mbox{\tiny{L}}}) = {\cal N}^{-1}\bm{M}_{fd}(y_{\mbox{\tiny{R}}},y_{\mbox{\tiny{L}}}){\cal N},
 \label{rel_Msv_Mfd}
\end{equation}
\noindent is obtained, which relates the two types of transfer matrices already defined.

Following the transfer matrix $\bm M_{sv}$ and scattering matrix $\bm S$ definitions we can establish
\begin{equation}\label{eq:perf-Msv}
 \begin{pmatrix}
 \overrightarrow{\bm\varphi}(y_{\mbox{\tiny{R}}})\\
 \overleftarrow{\bm \varphi}(y_{\mbox{\tiny{R}}})
 \end{pmatrix}=\bm M_{sv}\begin{pmatrix}
 \overrightarrow{\bm\varphi}(y_{\mbox{\tiny{L}}})\\
 \overleftarrow{\bm \varphi}(y_{\mbox{\tiny{L}}})
 \end{pmatrix}=\begin{pmatrix}
 \bm \alpha & \bm \beta\\
 \bm \gamma & \bm \delta
 \end{pmatrix}\begin{pmatrix}
 \overrightarrow{\bm\varphi}(y_{\mbox{\tiny{L}}})\\
 \overleftarrow{\bm \varphi}(y_{\mbox{\tiny{L}}})
 \end{pmatrix}
\end{equation}
and
\begin{equation}\label{eq:perf-S}
 \begin{pmatrix}
 \overleftarrow{\bm\varphi}(y_{\mbox{\tiny{L}}})\\
 \overrightarrow{\bm \varphi}(y_{\mbox{\tiny{R}}})
 \end{pmatrix}_{\textrm{out}}=\bm S\begin{pmatrix}
 \overleftarrow{\bm\varphi}(y_{\mbox{\tiny{L}}})\\
 \overrightarrow{\bm \varphi}(y_{\mbox{\tiny{R}}})
 \end{pmatrix}_{\textrm{in}}=\begin{pmatrix}
 \bm r & \bm t'\\
 \bm t & \bm r'
 \end{pmatrix}\begin{pmatrix}
 \overleftarrow{\bm\varphi}(y_{\mbox{\tiny{L}}})\\
 \overrightarrow{\bm \varphi}(y_{\mbox{\tiny{R}}})
 \end{pmatrix}_{\textrm{in}},
\end{equation}
respectively, where $\bm t$ and $\bm r$ ($\bm t'$ and $\bm r'$) are transmission and reflection amplitudes  for left (right) incident particles. Using (\ref{eq:perf-Msv}) and (\ref{eq:perf-S}) together with the imposed boundary conditions it is possible to relate the scattering amplitudes for the symplectic case\cite{Mello88,Leo06}
\begin{equation}\label{eq:AmpDisp}
 \begin{array}{lcl}
 \bm t=\bm \alpha- \bm \beta \bm \delta^{-1}\bm \gamma & \quad & \bm t'= \bm \delta^{-1}\\
 \bm r=-\bm \delta^{-1}\bm \gamma & \quad & \bm r'=\bm \beta \bm \delta^{-1}
 \end{array}.
\end{equation}
The connection between Transfer Matrix and Scattering Matrix formalisms is the most remarkable feature of MSA.

From relations (\ref{eq:AmpDisp}) we could obtain relevant transport magnitudes. Considering the incidence of particles from the left only, the transmission coefficient from channel $j$ to channel $i$ is given by
\begin{equation}
 \label{eq:T}
 T_{ij}=t_{ij}^*t_{ij},
\end{equation}
Furthermore, they allow to compute the conductance through the channel $i$
\begin{equation}
 \label{eq:Gb}
 G_{i}=\frac{e^2}{h}\sum_{j=1}^{N} t_{ij}^*t_{ij},
\end{equation}
the two-probe Landauer conductance
\begin{equation}
 \label{eq:GtpL}
 G=\frac{e^2}{h}\sum_{i=1}^{N}\sum_{j=1}^{N}T_{ij},
\end{equation}
the phase-transmission amplitudes
\begin{equation}
 \label{eq:Ph}
 \phi_{ij}=\arctan{\frac{Im\: t_{ij}}{Re\: t_{ij}}},
\end{equation}
and the phase-transmission times or group delays
\begin{equation}
 \label{eq:TPh}
 \tau_{ij}=\hbar\frac{\partial}{\partial E}\phi_{ij}.
\end{equation}

\section{Spin-dependent transport properties}
\label{sec:results}

Our \textit{gedanken} experiment consists of introducing spin-po\-la\-ri\-zed electrons in the three-region channel with SOC-R ($I$, $II$ and $III$ in Fig. \ref{fig:DDSFET} (b)), under normal incidence ($k_x=0$). We are able to tune up the coupling parameter $\alpha_{\mbox{\tiny{R}}}$ in the central region (region $II$, below the gate electrode), the length $L$ of this region and the energy $E$ of the incident electrons.
We set the length of the regions $I$ and $III$ as $l=10$~nm. SOC-R in these regions is solely given by the intrinsic asymmetry of the confining potential, which depends on the surface charge density, $n_s=10^{12}$~cm$^{-2}$, from now on.
Following the Landauer picture \cite{Imry99}, we assume the whole system connected to infinite reservoirs of charge at different chemical potential, avoiding the inclusion of an external electric field to move electrons along the system.

There are two direct paths connecting electrons with the same spin ($e_\uparrow\Rightarrow e_\uparrow$ and $e_\downarrow\Rightarrow e_\downarrow$) and two crossed paths connecting electrons with different spin ($e_\uparrow\Rightarrow e_\downarrow$ and $e_\downarrow\Rightarrow e_\uparrow$). The availability of the last two is given by the SOC-R. In accordance with our consideration for regions $F_{\mbox{\tiny{L}}}$ and $F_{\mbox{\tiny{R}}}$, we are only interested in the paths $e_\uparrow\Rightarrow e_\uparrow$ and $e_\uparrow\Rightarrow e_\downarrow$ because we are injecting in the channel only spin-$up$ polarized electrons.

\begin{figure}
\centering
\includegraphics[width=0.9\linewidth]{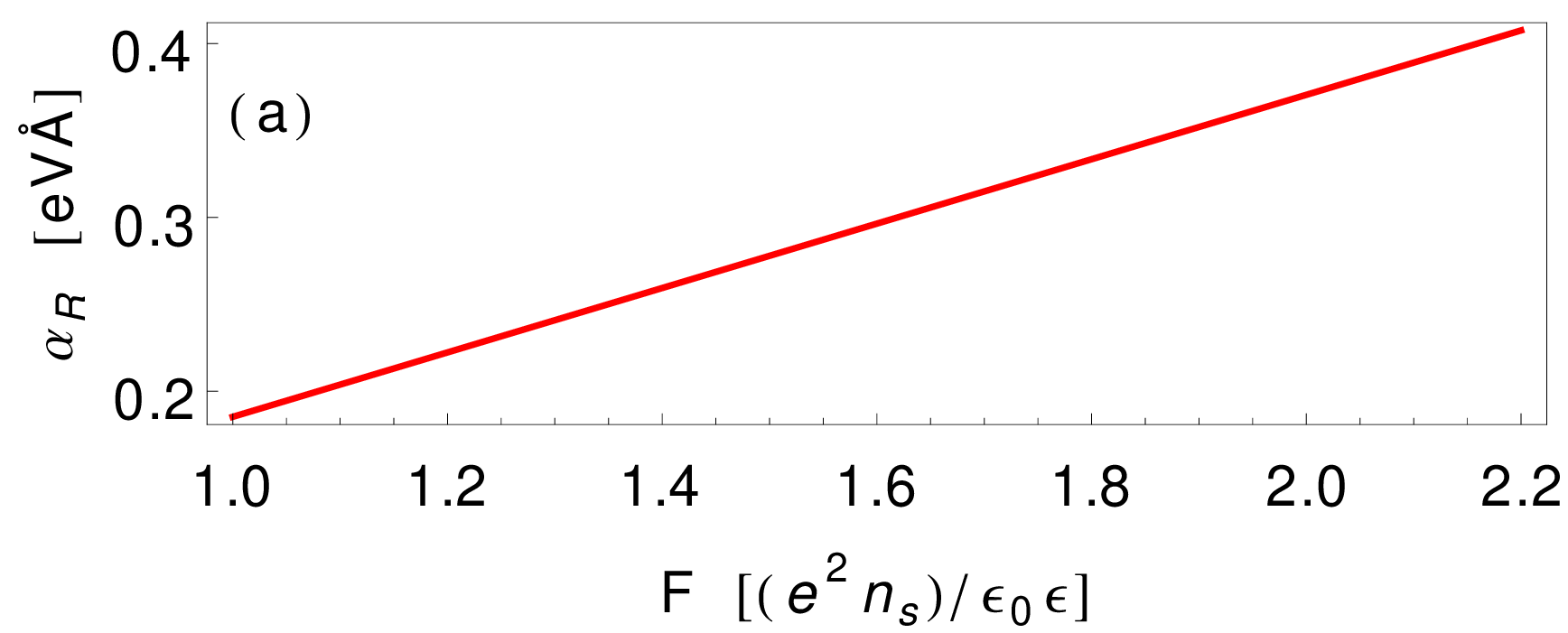}
\includegraphics[width=0.9\linewidth]{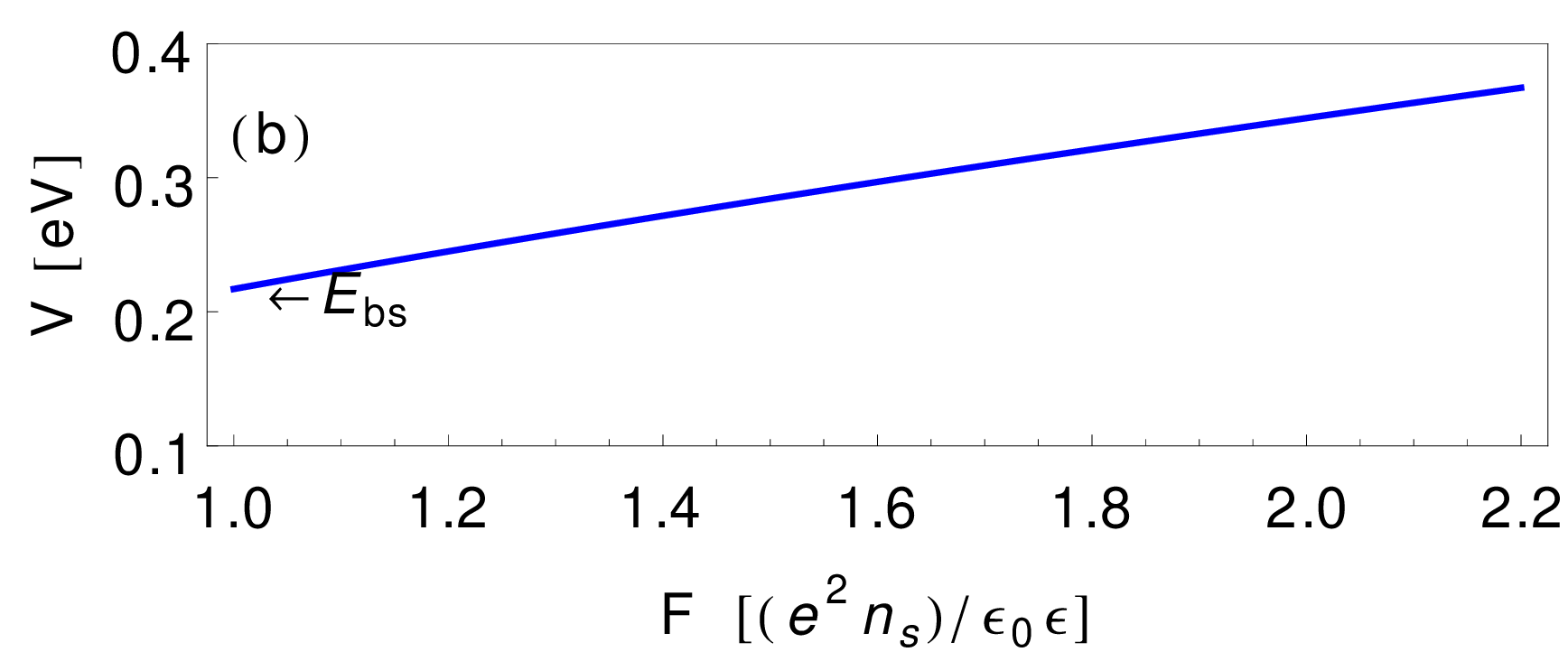}
\caption{(Color online) SOC-R coupling parameter $\alpha_R$ (\ref{eq:RashbaPar}) (panel (a)) and gate-voltage induced potential barrier $V$ (\ref{eq:Vpot}) (panel (b)) as function of the electric
 field at the region $II$, in units of the one at the bare regions ($I$ and $III$).}
\label{fig:AVenF}
\end{figure}

Taking as energy reference the first transversal state $E_{bs}$ ---given by Exp. (\ref{eq:AEz})--- electrons traveling from region $I$ into $II$ experience a potential barrier of height
\begin{equation}\label{eq:Vpot}
 V=C_1\left(\frac{\hbar^2}{2m^*}\right)^{1/3}\left(F_{II}^{\frac{3}{2}} -F_{I,III}^{\frac{3}{2}} \right),
\end{equation}
and thickness $L$, due to the mismatch of the energy levels between those regions, induced by the external electric field. Then, the potential $V$ essentially depends on the voltage at the gate electrode (region $II$, see Fig. \ref{fig:DDSFET}). Hence, by tunning $\alpha_{\mbox{\tiny{R}}}$ the height of this barrier also changes, as shown Fig.~\ref{fig:AVenF}. The coupling parameter $\alpha_{\mbox{\tiny{R}}}$ growths following (\ref{eq:RashbaPar}) (see panel (a)), while $V$ growths following (\ref{eq:Vpot}) (see panel (b)).

\begin{figure}
\centering
\includegraphics[width=0.9\linewidth]{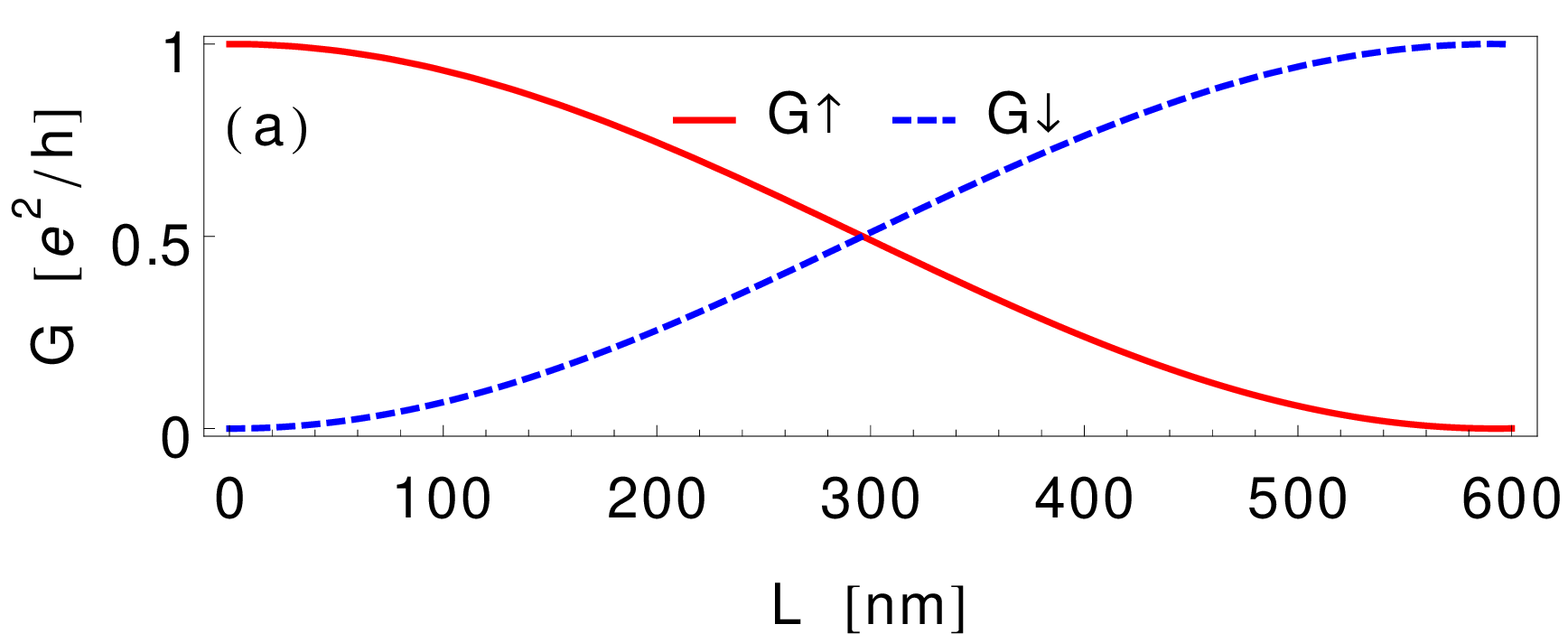}
\includegraphics[width=0.9\linewidth]{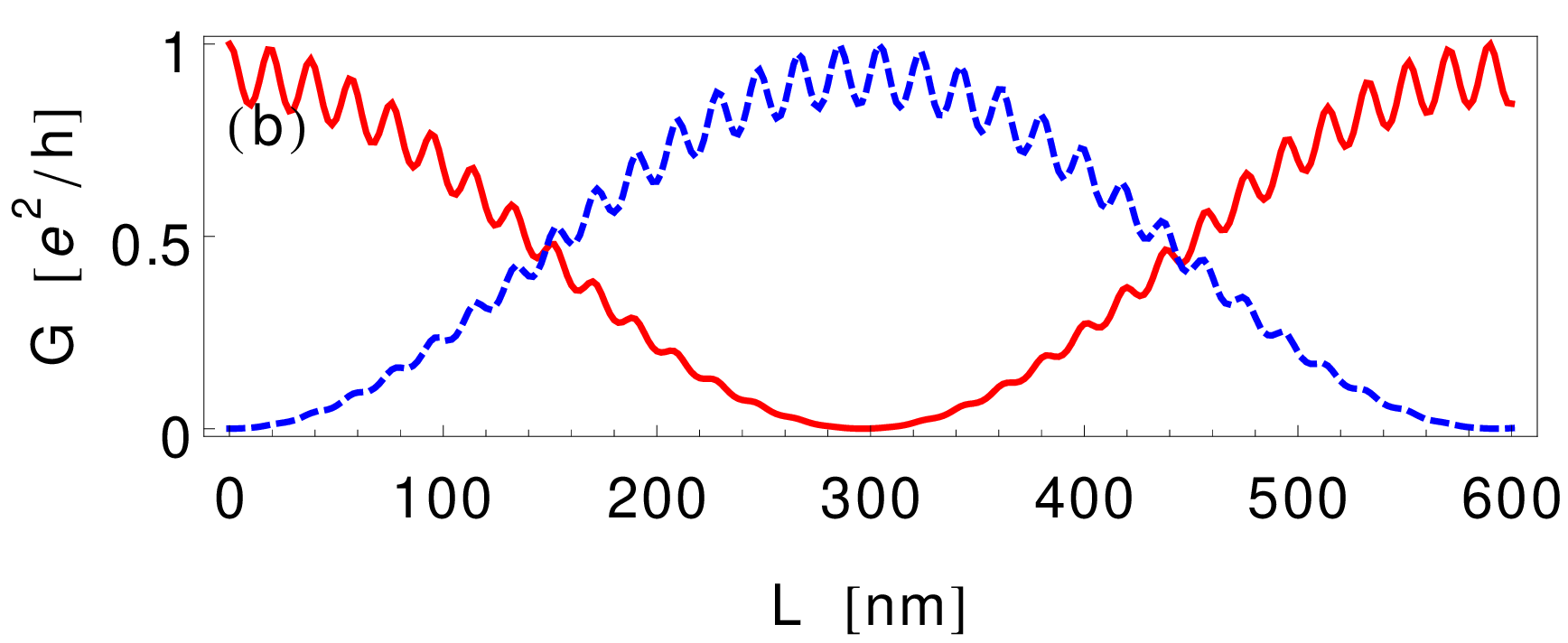}
\caption{(Color online) \textit{Closed} (a) and \textit{Opened} (b) SFET operating points: Fixing the polarization of incoming electrons as $\xi=|\uparrow\rangle_z$, a spin-$up$ $G_\uparrow$ or spin-$down$ $G_\downarrow$ polarized conductance is obtained at $L=600$~nm and $E=0.48$~eV for $F_{II}=F_{I,III}=\frac{e^2n_s}{\epsilon_0\epsilon}$ (panel (a) and $F_{II}=2.2F_{I,III}=2.2\frac{e^2n_s}{\epsilon_0\epsilon}$ (panel (b)).}
\label{fig:GenL}
\end{figure}

It is a well-known result \cite{Mireles01,Mireles02,Pala04} that spin-resolved conductance oscillates with the lengthening of the channel, due to the    SOC-R-induced spin precession phenomenon. In fact, assuming that electrons move freely (fill no potential barrier (\ref{eq:Vpot})) along the positive $y$ direction in the transport channel, following (\ref{eq:DispR})-(\ref{eq:Eigen}), the equivalent wave function would be
\begin{equation}\label{eq:toy}
 \bm\varPsi(y)=\frac{1}{\sqrt{2}}\begin{pmatrix}1 \\ i\end{pmatrix}e^{ik_{y1}y}+\frac{1}{\sqrt{2}}\begin{pmatrix}1 \\ -i\end{pmatrix}e^{ik_{y2}y},
\end{equation}
where $k_{y1}$ and $k_{y2}$ are the positive solutions of equations (\ref{eq:DispR}). Consequently, the probability of detecting spin-$up$ or spin-$down$ electrons at a given $L$ reads \cite{Mireles01}
\begin{equation}\label{ec:Tup}
 P_{up}=|\langle(1\:0)|\bm \Psi(L)\rangle|^2=\cos^2\left[ \dfrac{L}{2}(k_{y2}-k_{y1})\right],
\end{equation}
\begin{equation}\label{ec:Tdown}
 P_{down}=|\langle(0\:1)|\bm \Psi(L)\rangle|^2=\sin^2\left[ \dfrac{L}{2}(k_{y2}-k_{y1})\right],
\end{equation}
respectively. The oscillations are then straightforward from the presence of cosine and sine functions. This behavior is observed in our calculations, as a general trend (see Fig.~\ref{fig:GenL}).

Both SFET operating points are shown in the panels (a) and (b) of Fig.~\ref{fig:GenL}, respectively.   Fixing the polarization of incoming electrons along the $z$ direction $\xi=|\uparrow\rangle_z=\begin{pmatrix} 1\\0 \end{pmatrix}$ for convenience, at null gate-voltage the intrinsic asymmetry of the heterostructure potential ensures that spin-\textit{up} conductance $G_\uparrow$ (associated with the path $e_\uparrow\Rightarrow e_\uparrow$) vanishes for $L\cong600$~nm (Fig.~\ref{fig:GenL} (a), red solid line). Hence, by conservation low, the spin-\textit{down} polarized total outgoing flux maximizes ($G_\downarrow\approx 1$, Fig.~\ref{fig:GenL}~(a), blue dashed line) but being associated with the path $e_\uparrow\Rightarrow e_\downarrow$, becomes forbidden by the ferromagnetic analyzer, and then the device is \textit{closed}. On the other hand, for a suitable value of the gate-voltage that strengthens the electric field 2.2 times respect to the one at the bare regions ($F_{I,III}=\frac{e^2n_s}{\epsilon_0\epsilon}$), the conductance for spin-\textit{up} polarized electrons is magnified ($G_\uparrow\approx 1$ and $G_\downarrow\approx 0$), as shown in Fig.~\ref{fig:GenL} (b) (red solid and blue dashed lines, respectively). In this case, the ferromagnetic analyzer allows the transport, thereby the device is \textit{opened}. This two-states behavior of the SFET had been proposed \cite{Ladd10} for codifying and manipulating information (``$\emptyset$'' and ``$1$'' values of a bit) instead of the current field effect transistors.

\begin{figure}
\centering
\includegraphics[width=0.9\linewidth]{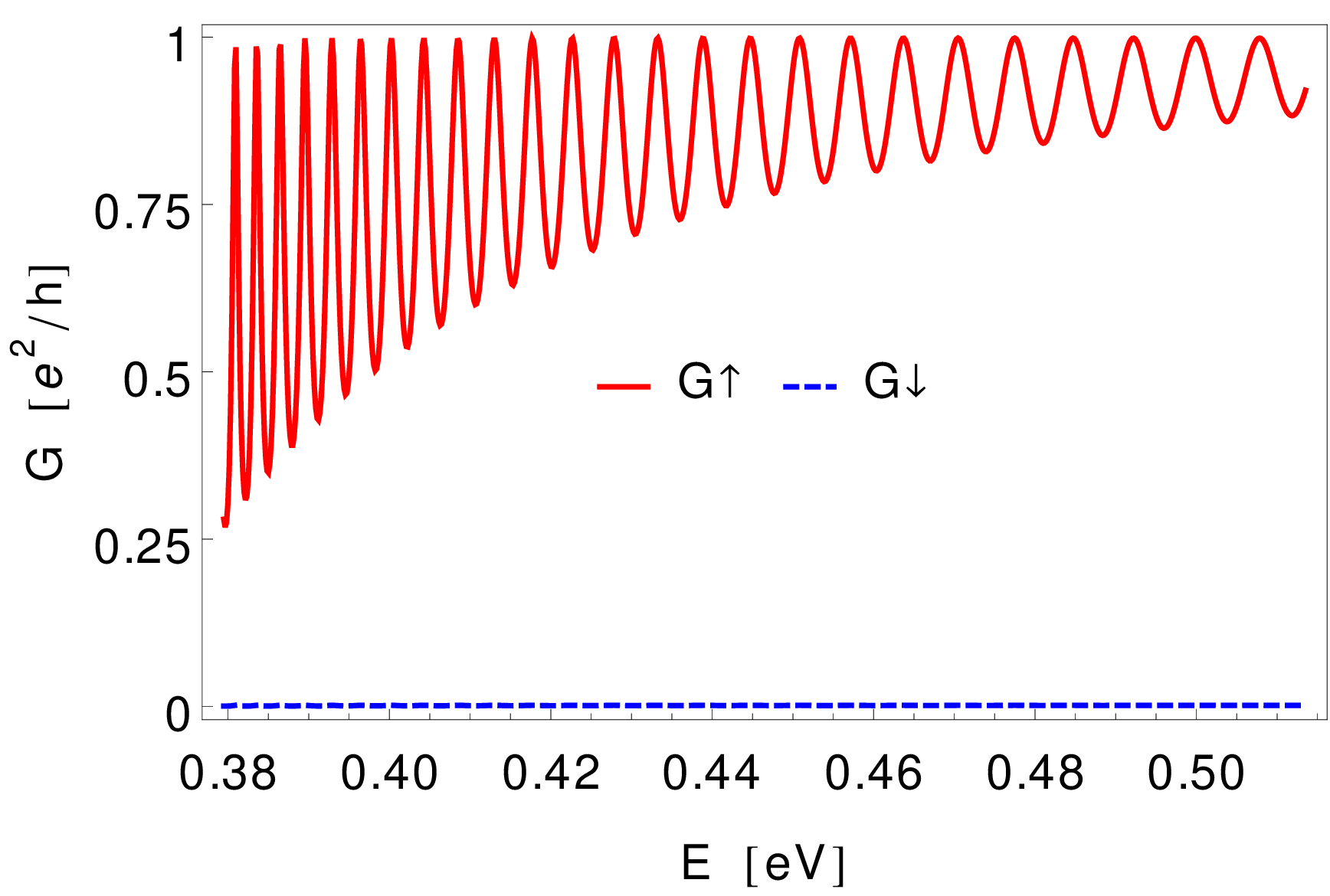}
\caption{(Color online) Spin-resolved conductance as function of the energy for the conditions represented in Fig.~\ref{fig:GenL} (b), at $L=600$~nm, $F_{II}=2.2F_{I,III}$ and $V(F_{II}) + E_{bs}=0.37$~eV.}
\label{fig:GenE}
\end{figure}

At a given value of the Fermi energy, by evaluating $k_{y1}$ and $k_{y2}$ in Exps. (\ref{eq:DispR}) we have $E_1(k_{y1})=E_2(k_{y2})$, then
\begin{equation}\label{eq:deltaK}
 \begin{split}
  E_2(k_{y2})-E_1(k_{y1}) & = 0,\\
  \dfrac{\hbar^{2}}{2m^{*}}(k_{y2}^{2}-k_{y1}^{2}) + \alpha_R(k_{y2}+k_{y1})& =0,\\
  \text{then } k_{y2}-k_{y1} & = -\dfrac{2m^{*}\alpha_R}{\hbar^{2}}.
 \end{split}
\end{equation}
From (\ref{eq:deltaK}) it is straightforward to note that $P_{up}$ (\ref{ec:Tup}) and $P_{down}$ (\ref{ec:Tdown}) do not depend on the energy \cite{Mireles01}, therefore the consequences of changing the energy on the spin-resolved conductance are typically leaved out. In our modelling, the energy plays a crucial role because we do consider the dispersive effects of the constriction at the region $II$ (see (\ref{eq:Vpot}) and its discussion). Actually,  the additional ``noise'' ---fast oscillations--- in spin-resolved conductance curves, at finite values of the gate-voltage (Fig.~\ref{fig:GenL} (b)), is clearly related with this. To evaluate the significance of those effects, in Fig.~\ref{fig:GenE} we plot the spin-resolved conductance  in a particular range of energy, setting all  parameters as in Fig.~\ref{fig:GenL} (b). The gate-voltage induced potential barrier (\ref{eq:Vpot}) has the height $V(F_{II}) + E_{bs} = 0.37 eV$, which is lower than the whole energy range considered, then in principle is not tunneling but transmission the quantum phenomenon we are considered. Note that the total conductance, $G_\uparrow+G_\downarrow$, could be reduced from 75\% down to 25\%. The existence of resonant peaks is a well-known phenomenon, associated to the condition $k(E)L=n\pi$, where $n$ is an integer.

\begin{figure}
\centering
\includegraphics[width=0.9\linewidth]{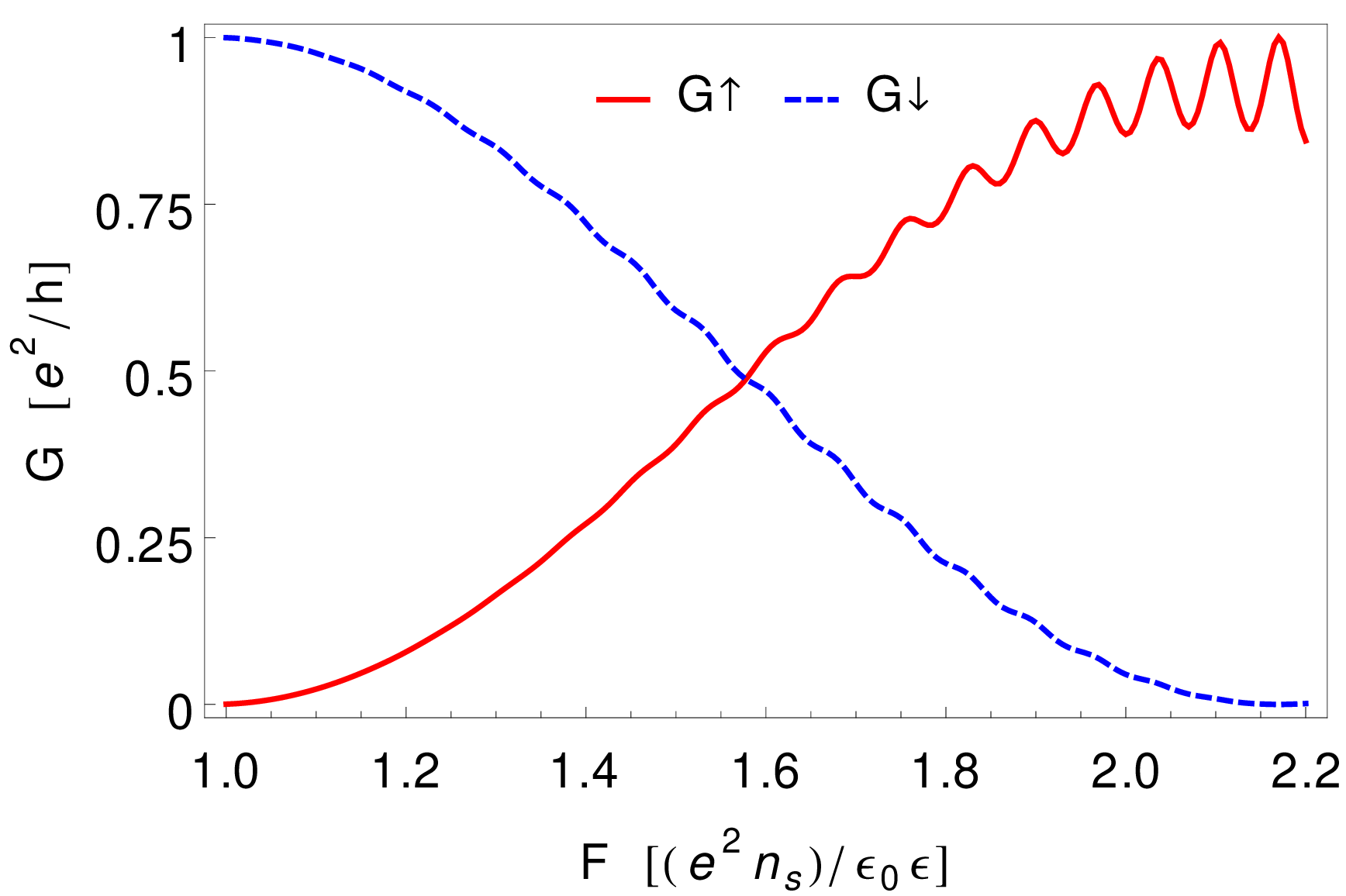}
\caption{(Color online) Spin-resolved conductance as function of the electric field at the region \textit{II} (induced by the gate electrode), in units of the one at the bare regions (\textit{I} and \textit{III}). $L=600$~nm and $E=0.48$~eV}
\label{fig:GenF}
\end{figure}

Fig.~\ref{fig:GenF} displays the spin-resolved conductance as function of the electric field at the region \textit{II} (induced by the gate electrode), in units of the one at the bare regions (\textit{I} and \textit{III}). This ($I-V$)-like curve characterizes the behavior of the modeled SFET. Note that for $F_{II}=F_{I,III}=e^2n_s/\epsilon_0\epsilon$ the outgoing flux is spin-$down$ polarized, leading to a \textit{closed} state, while for $F_{II}=2.2F_{I,III} = 2.2e^2n_s/\epsilon_0\epsilon$ is spin-$up$ polarized, leading to an \textit{opened} state. We chose the energy of the incoming spin-$up$ polarized electrons as $E=0.48$~eV, in order to avoid the zone of large-amplitude oscillations of the spin-resolved conductance (see Fig.~\ref{fig:GenE}). Even that, for the highest values of the electric field ($F_{II}\gtrsim 1.6F_{I,III}$) those oscillations are observed.

We are interested now in evaluating the quantum transport time in the system we are modelling (see Fig.~\ref{fig:DDSFET}). To answer how long it takes an electron to pass through a quantum system is still a controversial question \cite{Landauer94}. Phase time, defined as in (\ref{eq:TPh}), is a well-established definition of this time, and has been extensively used in researches regarding quantum tunneling \cite{Steinberg93,Winful03,Leo06,Leo07,Arias-Laso12}.

The first task then is to compute the phase as function of the energy following (\ref{eq:Ph}). Although not shown here, if one plots the phase  for the allowed paths, $\phi_\uparrow$ ($e_\uparrow\Rightarrow e_\uparrow$) and $\phi_\downarrow$ ($e_\uparrow\Rightarrow e_\downarrow$), one may note that the corresponding curves are shifted in the energy.  This is consistent with the fact that electrons with different spin polarization propagate along the system with different quasi-momentum, as in the ``toy model'' described by the solution (\ref{eq:toy}). Consequently, those curves have equal derivative and the phase time is the same along the direct and crossed paths. This may be explained taking into account that manipulation of spins in a SFET \textit{via} SOC-R ---in itself--- is not a dispersion-selective phenomenon thus the coherence is preserved. The opposite picture occurs in ferromagnets or dilute magnetic semiconductors, where spin separation features in time have been reported \cite{Guo01,Wang02}. A simple modelling of those systems \cite{Pala04,Yang08} predicts that they behave as a potential barrier for a particular spin polarization and as a potential well for the other, resembling an effective Zeeman splitting.

\begin{figure}
\centering
\includegraphics[width=0.9\linewidth]{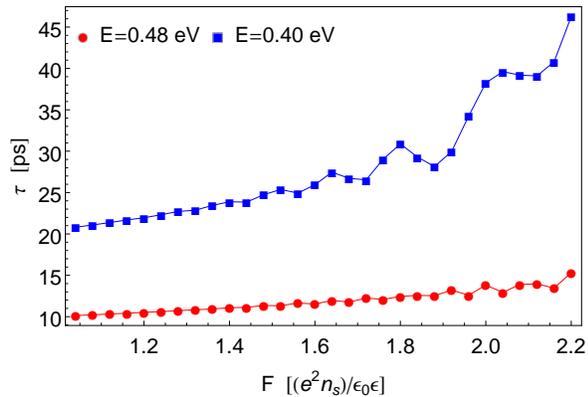}
\caption{(Color online) Phase time as function of the  electric field at the region \textit{II} for two particular values of the energy, $E=0.40$~eV (squares) and $E=0.48$~eV (circles). $L=600$~nm.}
\label{fig:TphenF}
\end{figure}

Fig.~\ref{fig:TphenF} displays the phase time $\tau$ as function of the  electric field at the region \textit{II} for two particular values of the energy, $E=0.40$~eV (squares) and $E=0.48$~eV (circles). At first sight, one may note that the phase time is lower for the higher energy and its value increases as a general trend in both cases, when one lets the electric field growths. This is intuitively understandable noting that the height of the potential barrier (\ref{eq:Vpot}) also growths, as shown in Fig.~\ref{fig:AVenF} (b). However, no significant changes are observed in $\tau$ for $E=0.48$~eV when one switches the device from the \textit{closed} state ($F_{II} = F_{I,III}$, depicted in Fig.~\ref{fig:GenL} (a)) to the \textit{opened} state ($F_{II} = 2.2F_{I,III}$, depicted in Fig.~\ref{fig:GenL} (b)). On the other hand, for $E=0.40$~eV, $\tau$ varies from $\sim 20$~ps to $\sim 45$~ps, which may be quite undesirable for practical applications. Another interesting issue here is the presence of oscillations in both curves, being more evident for $E=0.40$~eV. These oscillations have been reported before in the case of holes \cite{Leo07,Arias-Laso12}, but as a function of the energy for values above the potential barrier, therefore they have the same nature that the ones we are reporting here.

\begin{figure}
\centering
\includegraphics[width=0.9\linewidth]{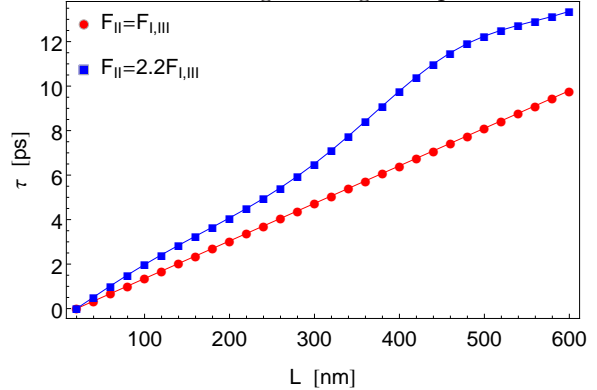}
\caption{(Color online) Phase time as function of the length $L$ of the region under SOC-R (region $II$) for both \textit{closed} ($F_{II}=F_{I,II}$, circles) and \textit{opened} (($F_{II}=2.2F_{I,II}$, squares)) states, with $E=0.48$~eV.}
\label{fig:TphenL}
\end{figure}

In Fig.~\ref{fig:TphenL} we plot $\tau$ as function of the length $L$ of the region under SOC-R (region $II$) for both \textit{closed} ($F_{II}=F_{I,II}$, circles) and \textit{opened} ($F_{II}=2.2F_{I,II}$, squares) states, with $E = 0.48$~eV. Note that for $L=600$~nm, which is a kind of spin-flip length for $n_s=10^{-12}$~cm$^{-2}$ (see Fig.~\ref{fig:GenL} (a)), the phase time for the \textit{opened} and \textit{closed} states barely differs in $\sim 4$~ps. While the phase time growths linearly with $L$ for $F_{II}=F_{I,II}$ (zero gate-voltage), exhibiting a classical-like behavior ($t_{f}=L\sqrt{m^{*}/(2E)}$), it follow a non-trivial increment when $F_{II}=2.2F_{I,II}$. From Figs.~\ref{fig:TphenF} and \ref{fig:TphenL} we may infer that characteristic times governing the operations of the device we are modelling are $\sim [10-50]$~ps, while the spin relaxation time is typically in the order of nanoseconds \cite{Balocchi11}.

It should be mentioned that is the first time that MSA formalism deal with systems of hundreds of nanometers, which is the typical scale length where spin-flip occurs \cite{Mireles01,Pala04,Yang08}. So far, the magnitudes (\ref{eq:T})-(\ref{eq:TPh}) had been successfully studied using MSA in systems with length ranging form $\sim 10$~\AA{} up to $\sim 100$~\AA{}. General properties, as the probability flux conservation through our system, was verified during the simulations. All physical parameters not mentioned above are referred to In$_{0.53}$Ga$_{0.47}$As  \cite{Vurgaftman01}.

\vspace{-5mm}

\section{Conclusions}
\label{sec:Conc}

The changes in the Datta-Das spin-transistor propagation medium opacity, given by the manipulation of the SOC-R strength, may considerably reduce the conductance of the device.  The spin-re\-sol\-ved conductance have exhibited oscillations as expected, but modulated by dispersive effects when the gate-voltage was turned on, leading to significant reduction of the total tunneling conductance. Nonetheless, in a zero temperature picture, present simulation allows us to identify suitable topological and external parameters, to obtain the operating points of the SFET and to prove its feasibility as substitute for current field effect transistors. Although we demonstrate that phase time may not be spin-resolved, we were able to evaluate the time that takes a spin-$up$ polarized electron to switch it down, resulting in a range of $\sim [10-50]$~ps, an order of magnitude lower than typical values of the spin relaxation times. The systematical procedure carried out here, provides the basis for studying novel SFET-like configurations \cite{Chuang15} and/or addressing $p$-doped SFET-like systems where the phenomenology is strikingly different respect to the electronic case \cite{BookWinkler03}. A work devoted to those issues is in progress and will be published elsewhere.

\bibliographystyle{aipnum4-1}
\bibliography{/home/rcuan/Documentos/Field-Work/Literatura/BibDB}

\end{document}